# Silicon nitride nanophotonics for low-redundancy 3D convolution

Shuai Hu[1,2], Jingcheng Li[1,2], Qichao Ding[3], Hongli Wang[3], Hailong Zhou[1,2,*], Chi Zhang[1,2,*], Jianji Dong[1,2,*] and Xinliang Zhang[1]

[1]*Wuhan National Laboratory for Optoelectronics, School of Optical and Electronic Information, Huazhong University of Science and Technology, Wuhan, China.*

[2]*Optics Valley Laboratory, 430074 Wuhan, China.*

[3]*Hubei Jiufengshan Laboratory, Wuhan, China.*

**Corresponding author: hailongzhou@hust.edu.cn; chizhang@hust.edu.cn; jjdong@hust.edu.cn;*

**Abstract:** Three-dimensional (3D) convolution extracts correlations in high-dimensional data, but overlapping receptive fields introduce substantial redundant data movement. Here, we demonstrate a silicon nitride 3D optical convolution accelerator (3D-OCA) that reconstructs receptive fields through coordinate-aware wavelength-to-group-delay mapping. A deterministically serialized input tensor is broadcast onto multiple wavelength channels, and a chirped waveguide Bragg grating (CWBG) compensates the temporal offsets associated with 3D kernel coordinates. This arrangement continuously forms neighboring receptive fields without repeatedly rearranging and loading their shared input samples. The integrated CWBG provides a differential group delay of 2691 ps and a dispersion of 124.8 ps/nm. At 20 Gbaud, spatial-spectral processing of Indian Pines data yields convolution agreement with a coefficient of determination up to 0.997 and 97.9% classification accuracy, compared with 98.9% digitally. At 10 Gbaud, the optical convolution layer preserves spatiotemporal features and achieves 92.5% accuracy on a four-class KTH video-recognition task. These results establish low-redundancy streaming 3D convolution across spectral and temporal data dimensions using the same optical delay architecture.

## Introduction

Three-dimensional (3D) convolutional neural networks (CNNs) extract local correlations jointly along three tensor dimensions[1,2], enabling spatiotemporal, spatial-spectral, and volumetric feature extraction[3]. They are widely used for high-dimensional data processing, including hyperspectral imaging and video analysis. However, their computational demand is accompanied by substantial data movement. Adjacent 3D convolution windows share many input elements, yet conventional implementations repeatedly access, rearrange, and reload these overlapping samples for multiply–accumulate operations. This redundancy motivates alternative architectures for efficient 3D tensor processing[4].

Optical computing exploits high bandwidth and parallelism through input broadcasting, weighting, delay, and photodetection[5]. Photonic architectures have demonstrated high-speed vector and image convolution{Citation}, but reconstructing a 3D receptive field from serialized data remains challenging. After deterministic serialization, neighbors along different tensor axes appear at different temporal separations, which cannot be described by a single uniform delay step[6]. A direct mapping from kernel-coordinate offsets to wavelength-dependent optical delays can recover these neighborhood relationships during propagation and reduce repeated electronic data preparation.

Here, we propose a chip-integrated native 3D optical convolution accelerator (3D-OCA) based on coordinate-aware wavelength-to-group-delay mapping. A single chirped waveguide Bragg grating (CWBG) compensates the temporal offsets introduced by tensor serialization. The input tensor is serialized once and broadcast onto wavelength channels, whose relative delays align samples belonging to the same receptive field. Continuous propagation then forms adjacent receptive fields without repeatedly rearranging and loading their shared samples. We demonstrate this low-redundancy streaming operation through Indian Pines hyperspectral classification[7] at 20 Gbaud and KTH video recognition at 10 Gbaud. The wavelength-delay channels and modulation rate can be programmed within the available spectral and delay ranges.

## Results

### Principle of the 3D-OCA

Three-dimensional convolution extracts local correlations along three tensor dimensions by applying a 3D convolution kernel over the input tensor. Let the input tensor be $X$, with dimensions $N_x \times N_y \times N_z$, where $N_x$, $N_y$, and $N_z$ represent the sizes along the three tensor dimensions, respectively. The 3D convolution kernel is denoted as $W$, with dimensions $K_x \times K_y \times K_z$, where $K_x$, $K_y$, and $K_z$ represent the kernel sizes along the three dimensions, respectively. A key challenge for optical 3D convolution is to reconstruct the samples belonging to the same local 3D receptive field after tensor serialization. In the proposed 3D-OCA (Fig. 1), the input 3D tensor is first serialized into a continuous temporal waveform according to a deterministic scan rule. The input tensor is serialized according to a deterministic scan order, with samples scanned first along the $x$ direction, then along $y$, and finally along $z$ Through serialization, the original 3D coordinate relationships are converted into deterministic temporal separations. The corresponding temporal offset relative to the first element is given by

$$\Delta \mathrm{t}(x, y, z) = \left[(x-1) + (y-1)N_x + (z-1)N_x N_y\right] T_s \quad (1)$$

where $x$, $y$, and $z$ denote the coordinates of the kernel element along the three tensor dimensions, $1 \leq x \leq K_x$, $1 \leq y \leq K_y$, and $1 \leq z \leq K_z$. $T_s$ represents the symbol interval between adjacent serialized samples. Therefore, the 3D coordinate relationships of convolution kernels are mapped into deterministic temporal offsets in the serialized data stream. For a $K_x \times K_y \times K_z$ convolution kernel, the maximum temporal offset required to cover the entire receptive field is given by

$$\Delta \mathrm{t}_{max} = \left[(K_x-1) + \left(K_y-1\right)N_x + (K_z-1)N_x N_y\right] T_s \quad (2)$$

This temporal aperture determines the delay range required for reconstructing a complete 3D receptive field.

The serialized waveform is subsequently broadcast onto multiple wavelength channels, with each wavelength corresponding to one position of the 3D convolution kernel. According to the coordinate-dependent temporal offsets, the wavelength channels are designed with corresponding group delays in the CWBG to compensate for the serialization-induced temporal separations. After propagation through the CWBG, samples that originally belong to the same 3D receptive field but are distributed at different temporal positions are re-aligned at the output. In this way, the convolution-kernel coordinate relationships are physically reconstructed through wavelength-dependent optical delays.

As the serialized data stream continuously propagates, adjacent 3D receptive fields are successively formed. Since neighboring convolution windows contain a large amount of overlapping data, repeated electronic data rearrangement and input loading for different receptive fields are avoided. Instead, the continuously serialized data stream, together with multi-wavelength broadcasting and wavelength-dependent group delays, progressively reconstructs adjacent 3D receptive fields over time, thereby enabling low-redundancy streaming 3D convolution. Different convolution configurations can further be supported by selecting different wavelength-delay channels within the available spectral and delay ranges.

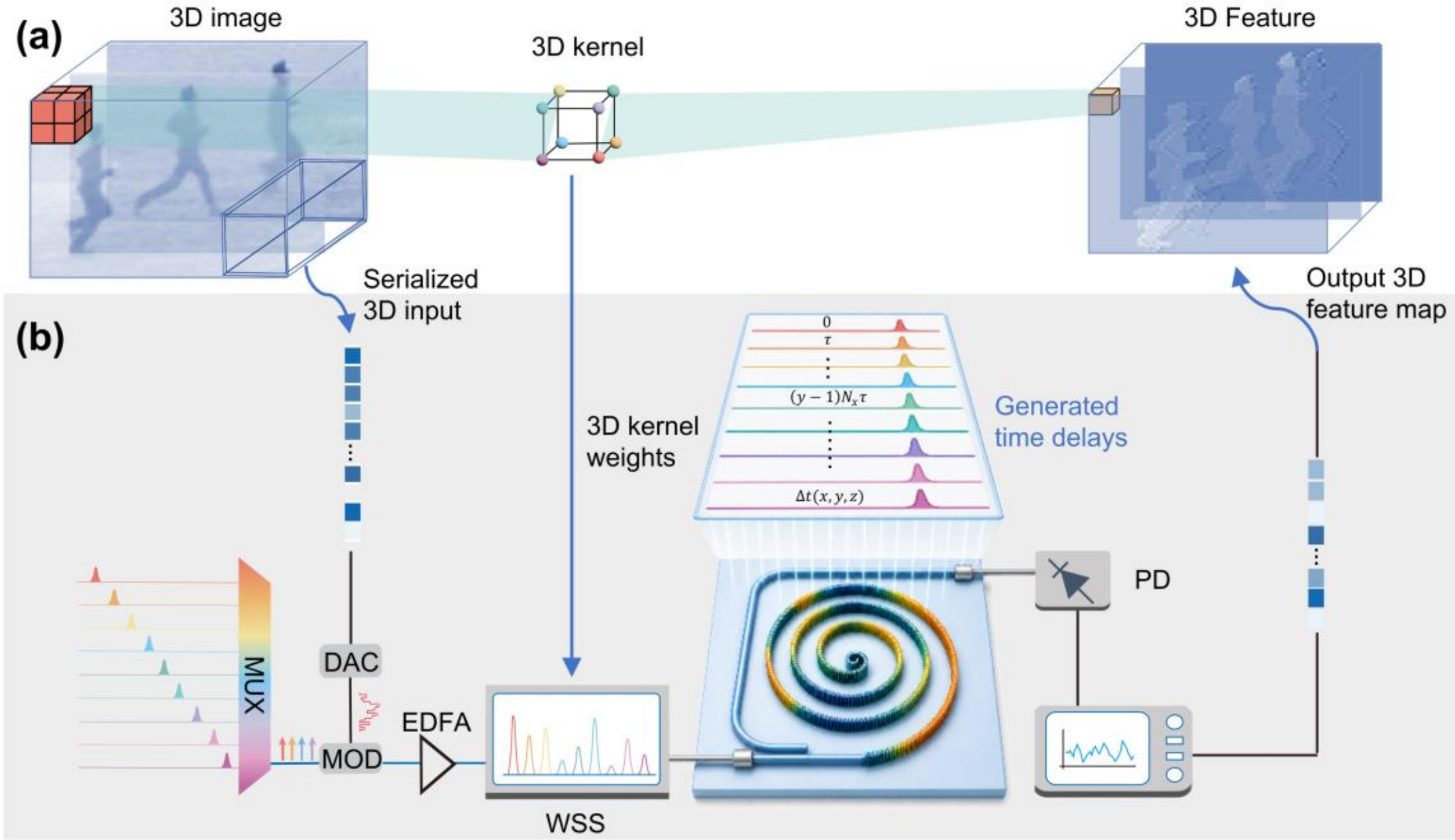

**Fig. 1. Principle and implementation of coordinate-aware wavelength-to-group-delay mapping for 3D optical convolution.** (a) A local 3D kernel extracts a feature tensor from the input. (b) The serialized waveform is broadcast onto wavelength channels, weighted, and delayed by the CWBG. The wavelength-dependent group delays compensate the coordinate-dependent temporal offsets, aligning samples within each receptive field for detection. Continuous propagation reconstructs adjacent receptive fields from the same input stream. MUX, multiplexer; DAC, digital-to-analog converter; MOD, modulator; EDFA, erbium-doped fiber amplifier; WSS, wavelength-selective switch; PD, photodetector.

The wavelength-dependent delays required for coordinate reconstruction are implemented using an integrated chirped waveguide Bragg grating (CWBG) fabricated on a silicon nitride platform[8]. The CWBG provides a designed wavelength-to-group-delay mapping, such that different optical carriers experience different group delays according to their assigned kernel coordinates[9]. The fabricated device provides a maximum differential group delay of approximately 2691 ps and a measured dispersion of approximately 124.8 ps/nm, offering sufficient temporal aperture for compensating the serialization-induced offsets of the selected 3D receptive fields.

In the experimental implementation, multiple wavelength carriers are combined and jointly modulated by the same serialized electrical waveform, allowing a single input stream to be broadcast across parallel optical channels. The wavelength channels are subsequently weighted by a wavelength-selective switch (WSS) and propagated through the CWBG, where the wavelength-dependent delays align samples belonging to the same receptive field. After delay compensation, the weighted optical channels are detected by a photodetector (PD), producing the serialized convolution output for subsequent feature reconstruction and classification.

## Spatial-spectral hyperspectral classification

We first apply the 3D-OCA to spatial-spectral data from a three-class subset of the Indian Pines hyperspectral dataset[10]. Corn-mintill, wheat and soybean-notill are selected because their partially overlapping spectral signatures provide a stringent test of joint spatial-spectral feature extraction. The tensor scan and channel delays are reprogrammed so that the third dimension represents spectral channels instead of a temporal coordinate, while the delay engine and weighted-detection mechanism remain unchanged. For the hyperspectral experiment, the serialized electrical waveform is operated at 20 Gbaud, corresponding to a 50-ps symbol interval, with each symbol encoding one spatial-spectral voxel. This experiment therefore tests whether wavelength-delay mapping generalizes across the physical meaning of the third data dimension.

The measured hyperspectral convolution outputs agree closely with their digital references, reaching a coefficient of determination of up to 0.997 across the experimental demonstrations. The waveform comparison, error distribution and correlation plot in Fig. 2 show that the spatial-spectral receptive field is reconstructed with high fidelity. The corresponding classification results demonstrate that the same optical primitive can process a third dimension that represents spectral correlation rather than time. The three-class hyperspectral classification achieves an overall accuracy of 97.9% with the optical convolution layer, compared with 98.9% for the matched digital reference. The small accuracy gap is consistent with the measured waveform-level deviations and indicates that wavelength-delay mapping preserves sufficient spatial-spectral feature information for hyperspectral classification.

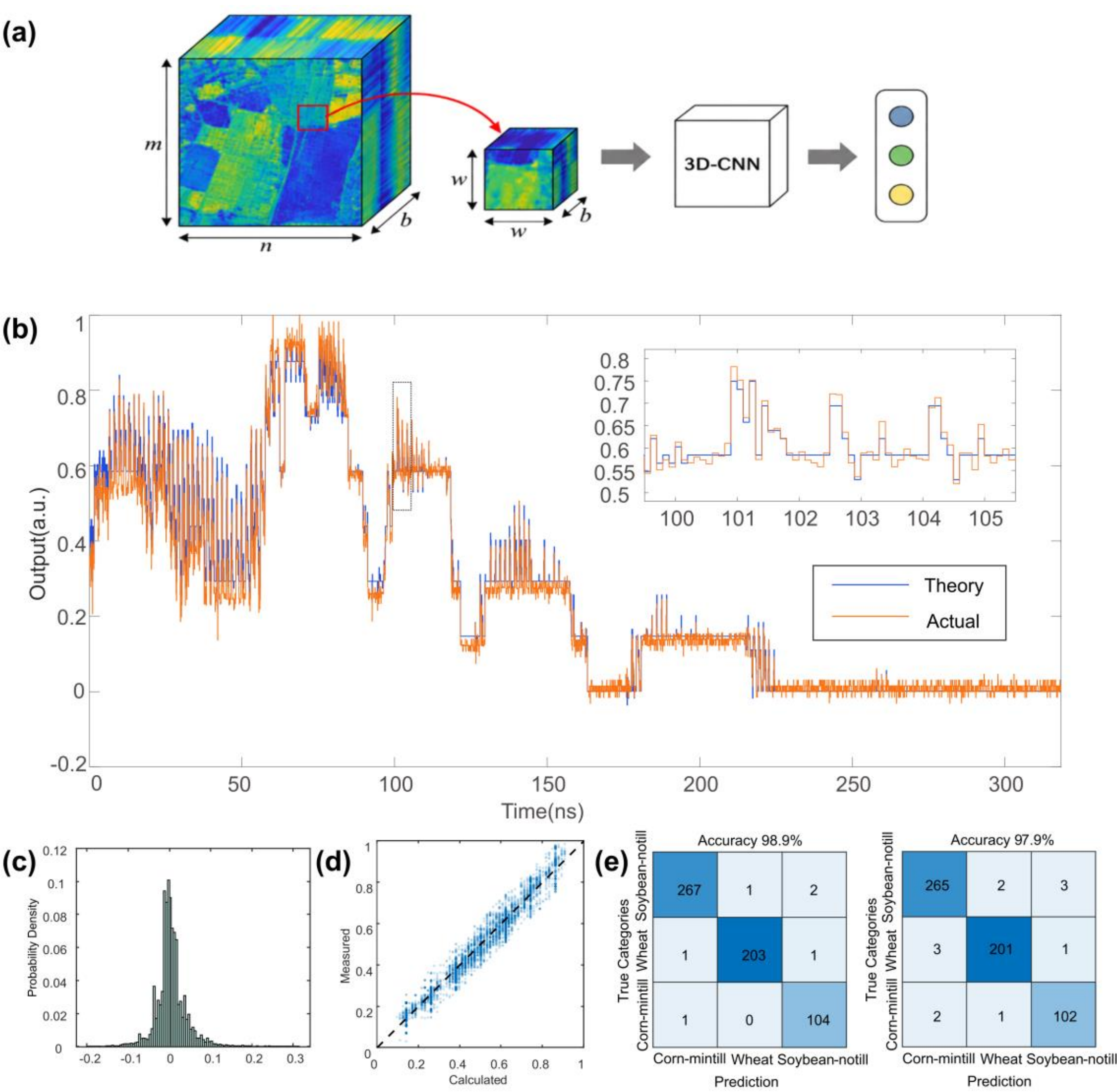

**Fig. 2. Indian Pines spatial-spectral convolution and classification.** (a) Hyperspectral processing workflow. (b) Measured and digitally calculated convolution waveforms, with a magnified segment. (c) Distribution of output deviations. (d) Correlation between measured and calculated outputs. (e) Confusion matrices for the digital and optical classifications.

The spatial-spectral convolution aggregates information across adjacent spectral bands while retaining local spatial context. The hyperspectral experiment therefore validates the generality of wavelength-delay mapping across different meanings of the third data dimension, rather than demonstrating only a task-specific optical filtering function.

## Spatiotemporal video recognition

We next evaluate whether the coordinate-aware wavelength-to-group-delay mapping preserves the spatiotemporal features required for video recognition[11,12]. A four-class subset of the KTH human-action dataset, comprising walking, running, boxing, and hand-waving, is processed using a hybrid optical–digital workflow: a digital 3D-CNN is trained offline, and the learned convolution weights are then programmed onto the optical weight engine (Fig. 3a). This protocol isolates the fidelity of the physical 3D convolution layer while maintaining a matched digital reference.

Each video block is serialized according to the mapping rule described above. The electrical waveform operates at 10 Gbaud, corresponding to a 100 ps symbol interval, with each symbol representing one voxel. The 3D kernel jointly samples local spatial structure and temporal evolution across adjacent frames. Accurate reconstruction requires the relative CWBG delays to match the temporal offsets of the serialized voxels, directly testing the delay-alignment fidelity of the optical engine.

Measured optical waveforms closely follow the corresponding digital convolution waveforms (Fig. 3b). The magnified segment in Fig. 3c resolves agreement in the local waveform transitions and amplitudes, supporting the temporal alignment required to reconstruct the spatiotemporal receptive fields. After inverse mapping, the recovered feature maps retain the motion trajectories and dynamic edge evolution of the input actions (Figs. 3f–3i). The root-mean-square errors (RMSEs) between the experimental and simulated feature maps are 0.0224, 0.0328, 0.0257, and 0.0274 for Figs. 3f–3i, respectively. These values quantify the agreement between optical and digital feature reconstruction and connect waveform fidelity to task-relevant spatiotemporal representations.

The optical CNN achieves an overall classification accuracy of 92.5% on the four-class KTH task, compared with 95.3% for the matched digital reference (Figs. 3d and 3e). Residual differences are associated with similar motion patterns and experimental non-idealities, including polarization drift, laser noise, and inter-channel gain imbalance. These results show that the optical layer preserves sufficient spatiotemporal feature information for recognition. Overlapping input samples are reused through the continuous stream, avoiding repeated electronic rearrangement and loading for each receptive field.

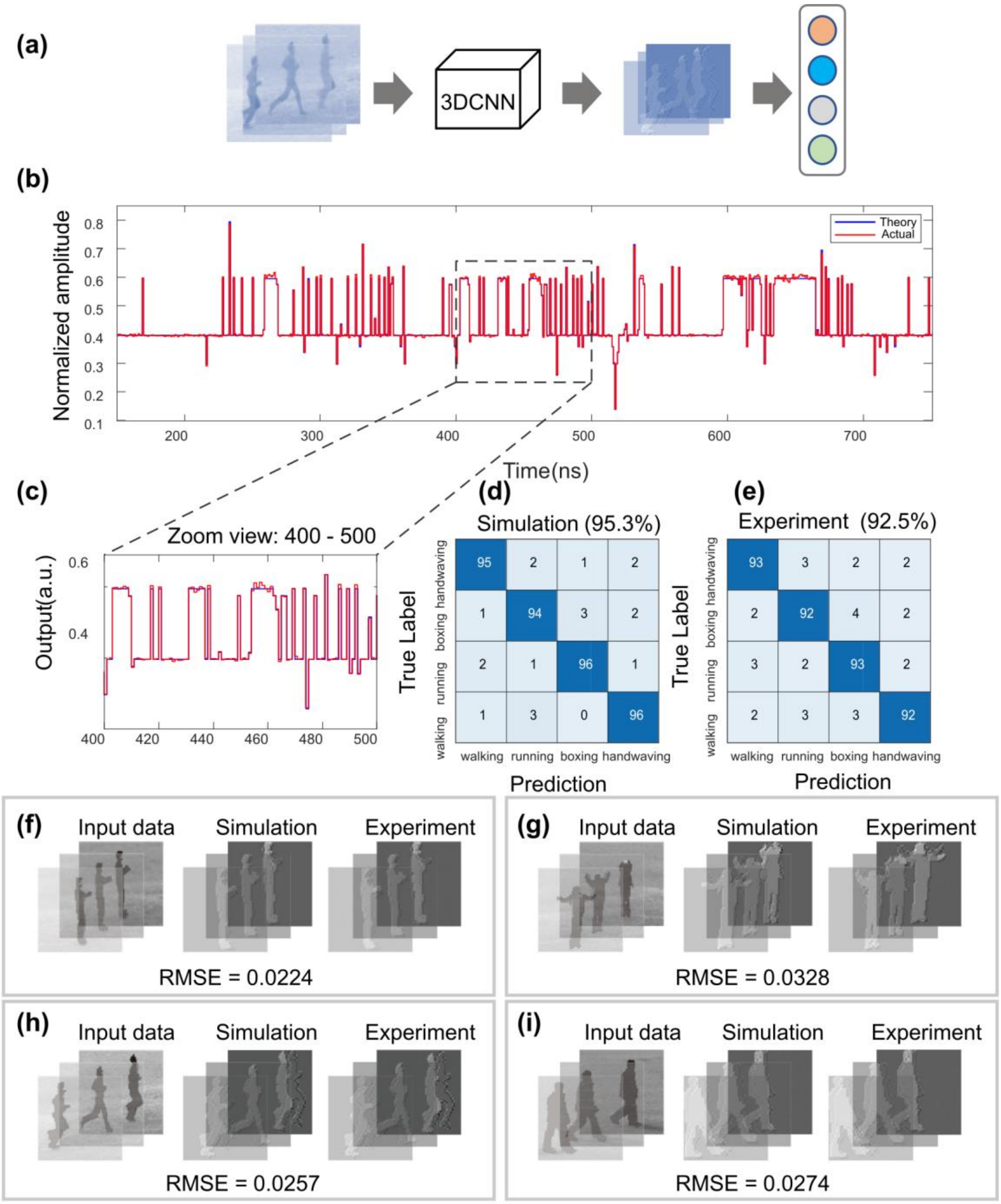

**Fig. 3. Spatiotemporal 3D convolution and video recognition on the KTH dataset.** (a) Hybrid optical–digital recognition workflow. (b) Measured and digitally calculated convolution waveforms. (c) Magnified waveform comparison. (d,e) Confusion matrices for the digital simulation (95.3%) and optical experiment (92.5%). (f–i) Input video data and corresponding simulated and experimental feature reconstructions, with root-mean-square errors (RMSEs).

## Conclusion

We have demonstrated a silicon nitride 3D-OCA based on coordinate-aware wavelength-to-group-delay mapping. The CWBG compensates serialization-induced temporal offsets to reconstruct adjacent 3D receptive fields from a continuous input stream, avoiding repeated electronic rearrangement and loading of overlapping samples. The optical layer achieves 97.9% Indian Pines classification accuracy at 20 Gbaud, compared with 98.9% digitally, and 92.5% KTH video-recognition accuracy at 10 Gbaud. These results demonstrate low-redundancy streaming 3D convolution for both spatial-spectral and spatiotemporal data using the same optical delay architecture.